# Cryogenic Magnetization Dynamics in Chemically Stabilized, Tensile-Strained Ultrathin Yttrium Iron Garnets with Tunable Magnetic Anisotropy

Jihyung Kim[1†], Dongchang Kim[2†], Seung-Gi Lee[1†], Yung-Cheng Li[2], Jae-Chun Jeon[2], Jiho Yoon[2], Sachio Komori[3], Ryotaro Arakawa[3], Tomoyasu Taniyama[3], Stuart S. P. Parkin[2*], and Kun-Rok Jeon[1*]

[1]*Department of Physics, Chung-Ang University (CAU), 06974 Seoul, Republic of Korea*

[2]*Max Planck Institute of Microstructure Physics, Weinberg 2, 06120 Halle (Saale), Germany*

[3]*Department of Physics, Nagoya University, Nagoya 464-8602, Japan*

[†]These authors contributed equally to this work.

*To whom correspondence should be addressed: stuart.parkin@halle-mpi.mpg.de, jeonkunrok@gmail.com

**ABSTRACT**

**We report an interfacial chemical stability–driven reduction of low-temperature damping losses in tensile-strained, ultrathin $Y_3Fe_5O_{12}$ (YIG) films grown by pulsed laser deposition, exhibiting ultralow room-temperature damping constants and tunable magnetic anisotropy. Comparative broadband FMR measurements show that tensile-strained YIG films on $Gd_3Sc_2Ga_3O_{12}$ (GSGG) retain measurable damping even at nanometer thicknesses and cryogenic temperatures down to 2 K, outperforming relaxed films on $Gd_3Ga_5O_{12}$. Based on static magnetometry measurements along with microstructural and compositional analyses, we attribute these enhanced dynamic properties to the suppression of interdiffusion across the YIG/GSGG interface, resulting from enhanced chemical stability and favorable growth kinetics by the presence of Sc. Our findings**

**highlight the importance of chemical and kinetic factors in achieving few-nanometer-thick YIG film with negligible low-temperature damping dissipation and perpendicular magnetic anisotropy for cryogenic spintronic applications.**

## 1. INTRODUCTION

The field of magnon spintronics fundamentally relies on magnetic insulators with low Gilbert damping, which enable the efficient generation, propagation, and control of spin waves[1-3]. Among these materials, yttrium iron garnet $Y_3Fe_5O_{12}$ (YIG) continues to serve as the reference standard, attributed to its exceptionally low magnetic damping[4], long magnon diffusion length[2], and excellent compatibility with microwave-frequency technologies[1-3]. Hence, YIG has played a central role in the development of magnonic and hybrid quantum systems[5-8], including resonant spintronic elements[9], spin-wave logic devices[10], and cavity magnon-polariton platforms[11,12].

With the growing demand for nanoscale and cryogenically operable spintronic devices[13], new material challenges have emerged. A central issue is maintaining ultralow magnetic damping in YIG films as their thickness is reduced to just a few nanometers and as device operation extends into cryogenic temperature regimes, where magnetization dynamics become highly sensitive to defects and impurities. While bulk and thick YIG films grown by liquid phase epitaxy[14,15] can exhibit remarkably low Gilbert damping constants ($\alpha \sim 10^{-5}$), this desirable property deteriorates significantly in thin films[16,17], primarily due to enhanced surface and interface scattering[18], the presence of impurities and defects[16,18], and interdiffusion across the film/substrate interface[19,20].

Recent studies employing thin-film deposition techniques, for instance, pulsed laser deposition (PLD)[17,21-23] and sputtering[24-26], have achieved the epitaxial growth of high-quality, ultrathin YIG films on garnet substrates. However, realizing low Gilbert damping and stable perpendicular magnetic anisotropy (PMA) in these films remains strongly dependent on the choice of substrate, as well as precise control of growth parameters and post-growth annealing. While earlier studies[17,21-23] predominantly used gadolinium gallium garnet $Gd_3Ga_5O_{12}$ (GGG) substrates due to their close lattice match with YIG, recent research[24-26] suggests that gadolinium scandium gallium garnet $Gd_3Sc_2Ga_3O_{12}$ (GSGG), which induces largest epitaxial tensile strain in the YIG layer among typical garnet substrates, offers a promising alternative.

Although substrate-induced strain is a well-established method for tuning magnetoelastic anisotropy in thin films, the specific influence of chemical composition and growth kinetics, particularly those related to the substrate, on the magnetic damping of ultrathin (< 20 nm) YIG films at cryogenic temperatures remains largely unexplored. This ultrathin YIG regime is particularly relevant for cryogenic magnonic and spintronic platforms[5–12], where strong spin-wave confinement, interfacial and proximity effects in heterostructures, and compatibility with nanoscale or quantum-hybrid circuits favor magnetic layers only a few nanometers. The emergence of PMA in these ultrathin films critically impacts spin-wave dynamics at both room and cryogenic temperatures by modifying dispersion, altering mode confinement and excitation efficiency, and enabling higher device densities in nanoscale magnonic architectures[1]. In combination with a heavy metal, PMA facilitates fast magnetization dynamics via spin-orbit-torque switching[27] and supports high domain-wall velocities in magnetic-insulator-based racetrack devices[28,29].

In this work, we present a detailed comparative study of ultrathin (3–10 nm) YIG films grown on two different substrates: lattice-matched GGG and lattice-mismatched GSGG. Our investigation reveals pronounced differences in both dynamic and static magnetic properties, including magnetic damping losses, magnetic anisotropy, and saturation magnetization. These findings highlight the substantial influence of chemical stability and growth kinetics over substrate-induced strain on the YIG's magnetic properties, which ultimately govern the performance and suitability of YIG films for cryogenic spintronic and magnonic applications,

## 2. EXPERIMENTAL SECTION

Ultrathin YIG films are pulsed-laser deposited onto single-crystal GGG(111) and GSGG(111) substrates, each with lateral dimensions of 10 mm × 10 mm, in an ultra-high vacuum chamber with a base pressure below $1\times10^{-7}$ Torr. A KrF excimer laser ($\lambda$ = 248 nm) is operated at a repetition rate of 10 Hz with an energy density of 1 $J/cm^2$ to ablate a stoichiometric YIG target. During deposition, the substrate temperature is maintained at 750 °C, and the oxygen partial pressure is controlled at 300 mTorr to ensure adequate oxygen incorporation and stoichiometry in the film.

The YIG film thickness $t_{YIG}$ ranges from 3 to 10 nm, as confirmed by X-ray reflectivity (XRR, Supplementary Section 1). Within this ultrathin regime, epitaxial tensile strain, originating from lattice mismatch between the film and the substrate, plays a crucial role in modifying the magnetic anisotropy[24-26]. Given the lattice constant of bulk YIG ($a_{YIG}$ = 12.376 Å in the (111) at room temperature), the GSGG substrate, with its larger lattice constant ($a_{GSGG}$ = 12.572 Å) compared to the GGG ($a_{GGG}$ = 12.383 Å), induces a stronger tensile strain in the YIG films. This, in turn, promotes perpendicular magnetic anisotropy (PMA) and influencing the magnetization orientation and effective damping[24-26]. The epitaxial quality and strain state of the films are confirmed by high-resolution X-ray diffraction (HRXRD) measurements performed using a Bruker D8 Discover diffractometer with Cu Kα radiation ($\lambda$ = 1.5406 Å). Surface morphology and roughness are studied using a Bruker Dimension Icon atomic force microscope (AFM).

We characterize the static magnetization properties of the same samples at 300 K using a Lake Shore vibrating sample magnetometer (VSM). In these static measurements, the magnetic field $\mu_0 H$ is applied either parallel or perpendicular to the film plane to estimate magnetic anisotropy and saturation magnetization.

The ferromagnetic resonance (FMR) response of the sample, mounted on a broadband coplanar waveguide, is investigated using DC field modulation techniques[30,31]. To acquire each FMR spectrum, we measure the microwave (MW) power absorbed by the sample while sweeping the external magnetic field $\mu_0 H$ at the fixed MW frequencies ranging from 0.5 to 20 GHz. For all FMR measurements, the MW input power is set to 10 dBm, corresponding to a few milliwatts of actual power absorbed by the sample. The magnetic field $\mu_0 H$ is applied parallel to the film plane. At the start of each measurement, we apply a large in-plane (IP) field ($\mu_0 H > 0.3$ T) to fully magnetize the YIG layer and then reduce the field to the FMR range. We use a Quantum Design Physical Property Measurement System (PPMS), which allows the application of $\mu_0 H$ up to 9 T over a temperature $T$ range of 2–400 K.

## 3. RESULTS AND DISCUSSION

**3.1. Relaxed and tensile-strained epitaxial ultrathin YIG films.** Figures 1(a)–(d) display the $\theta$–$2\theta$ XRD scans and XRR profiles of ultrathin YIG films ($t_{YIG}$ = 3–10 nm), grown on either GGG(111) or GSGG(111) substrates. The presence of (444) diffraction peaks in Figs. 1(a) and 1(c) confirms the epitaxial nature of ultrathin YIG films for all thicknesses on both GGG(111) and GSGG(111) substrates. Compared to YIG films grown on lattice-matched GGG substrates [inset of Fig. 1(a)], those grown on lattice-mismatched GSGG substrates [inset of Fig. 1(b)] exhibit a shift of the (444) diffraction peak to a higher $2\theta$ angle. This shift indicates a reduction in the out-of-plane (OOP) lattice constant, which results from IP tensile strain induced by the lattice-mismatched substrate. To maintain the overall unit cell volume, the IP expansion caused by this tensile strain is compensated by a contraction in the OOP direction[24–26].

The magnetic anisotropy energy $E_K$ for ultrathin YIG films, primarily governed by the magnetic shape anisotropy $K_s$ and the uniaxial magnetoelastic anisotropy $K_u$, both of which dominate over the cubic magnetocrystalline anisotropy $K_c$, is approximated by[32]:

$$E_K \approx (K_s + K_u) sin^2(\theta). \qquad (1)$$

Here, $K_s = -\frac{\mu_0}{2} M_s^2$ and $K_u = \frac{3}{2}\lambda_{111}\sigma_{111}$, where $\lambda_{111}$ is the magnetostriction constant along the (111) direction. $\sigma_{111} = \frac{Y}{1+\nu}\varepsilon_\perp$ is the uniaxial stress along the (111) direction with the OOP misfit $\varepsilon_\perp$. $Y$ and $\nu$ represent Young's modulus and Poisson's ratio, respectively. The OOP strain is elastically related to the IP misfit as $\varepsilon_\perp = -1.296\varepsilon_\parallel$, where $\varepsilon_\parallel = \frac{a_{sub} - a_{film}}{a_{film}}$ defines the IP misfit and $a_{film}$ ($a_{sub}$) denotes the lattice constant of the film (substrate). The angle $\theta$ represents the orientation of the magnetization relative to the YIG(111) direction. For YIG[32-34], the parameters are $\lambda_{111} = -2.4\times10^{-6}$, $Y = 2\times10^{12}$ Nm$^{-2}$, and $\nu = 0.29$. Using these values and for the GSGG substrate, we obtain $\varepsilon_\parallel = +0.014$ (IP tensile strain) and $\varepsilon_\perp = -0.0187$ (OOP compressive strain). Consequently, in the ultrathin limit, $K_u = +104$ kJm$^{-3}$ exceeds $K_s = -12.3$ kJm$^{-3}$, resulting in a positive value of $K_s + K_u$, which energetically favors magnetization along the (111) axis, perpendicular to the film plane, thereby inducing PMA. However, as the YIG thickness increases and the strain relaxes, $\sigma_{111}$ decreases, leading to a reduction in $K_u$ and thus a weakening of the PMA.

All the relaxed and tensile-strained YIG films exhibit atomically smooth surfaces [Figs. 1(c) and 1(d)], with a root-mean-square (RMS) roughness ranging from 0.12 to 0.16 nm and a typical standard deviation of 0.02–0.03 nm.

**3.2. Tensile strain–promoted perpendicular magnetic anisotropy.** The aforementioned

PMA in tensile-strained ultrathin YIG films is probed using static magnetometry measurements performed with a VSM. Figure 2(a) shows the measured IP and OOP magnetization $\mu_0 M$, normalized by their respective saturation magnetization $\mu_0 M_s$ values, versus $\mu_0 H$ for YIG films ($t_{YIG}$ = 3–10 nm) grown on either GGG(111) or GSGG(111) substrates. For the YIG films on GGG(111) substrates, the magnetization easy axis remains predominantly IP, exhibiting in-plane magnetic anisotropy (IMA), down to $t_{YIG}$ = 3 nm [Fig. 2(a)], consistent with prior studies[17,22,23]. In contrast, for the films on GSGG(111) substrates, the easy axis undergoes a transition from IP to OOP orientation, *i.e.,* from IMA to PMA, as $t_{YIG}$ decreases from 10 nm to 3 nm [Fig. 2(b)]. These results indicate that the observed PMA arises from interfacial effects and is qualitatively consistent with uniaxial magnetoelastic anisotropy along the YIG(111) direction, induced by IP tensile strain (or, OOP compressive strain) from the substrate, as discussed above and in Ref. [26]. We note that for $t_{YIG}$ = 10 nm with $\varepsilon_{\perp} \approx -0.002$ [estimated from Fig. 1(c)], $K_u$ = +11 kJm$^{-3}$ is smaller than $K_s$ = −12.3 kJm$^{-3}$, leading to a negative value of $K_s + K_u$ and thus the dominance of IMA over PMA. This quantitative analysis accounts for the observed IMA-to-PMA transition as $t_{YIG}$ decreases from 10 to 3 nm.

From the areal saturation magnetization $M_s \cdot t_{YIG}$ versus $t_{YIG}$ plots in Fig 2(c), the effective thickness of the magnetic dead layer $\delta$, *i.e.,* an interfacial region with zero net magnetic moment at the YIG/substrate interface, can be estimated by extrapolating the intercept of a linear fit to the $M_s \cdot t_{YIG}$ data, using the following relationship,

$$M_s = M_0 \left(1 - \frac{\delta}{t_{YIG}}\right), \qquad (2)$$

where $M_0$ is the slope of the linear fit and relates to the volume saturation magnetization[35]. Notably, we estimate the value of $\delta$ to be approximately 2.3 nm for YIG films grown on GGG(111) substrates and about 0.7 nm for those grown on GSGG(111) substrates, suggesting that the formation of the magnetic dead layer is largely suppressed when YIG is grown on GSGG(111). This is, somehow, contrary to what would be expected based on thermodynamic considerations. Typically, lattice-misfit-induced strain and dislocations enhance interdiffusion, potentially leading to the formation of a more pronounced magnetic dead layer[36], provided that the atomic species are mobile and the interface chemistry is conducive.

**3.3. Enhanced chemical stability by the presence of Sc.** This anomalous behavior can be attributed to the presence of scandium ($Sc^{3+}$) in the GSGG substrate. Compared to gallium ($Ga^{3+}$), scandium ($Sc^{3+}$) is a harder cation with higher field strength and forms stronger ionic bonds within garnet lattices, despite its slightly larger ionic radius[37–39]. Owing to its relatively low mobility in garnet structures, $Sc^{3+}$ is unlikely to readily substitute for $Fe^{3+}$ sites in YIG. In addition, even under identical growth conditions, subtle differences in surface termination (e.g., Sc-terminated versus Ga-terminated surfaces) and surface diffusion barriers can lead to distinct nucleation behavior and reduced intermixing during the early stages of YIG growth.

The suppression of Fe/Ga interdiffusion upon Sc substitution can be semi-quantitatively understood in terms of both ionic size mismatch and bonding energetics. Substitution of $Ga^{3+}$ by $Sc^{3+}$ introduces local lattice distortions and stronger cation–oxygen interactions, which increase the energetic cost of cation migration across the interface. In particular, the higher field strength and stronger Sc–O bonding character compared to Ga–O increase the migration energy barrier $E_m$ for diffusion[37–39], leading to an exponential reduction of interdiffusion according to Arrhenius kinetics, $D \propto \exp(-E_m/k_B T)$, where $D$ is the diffusion coefficient and $k_B$ is the Boltzmann constant. The resulting chemically sharper interface suppresses Fe–O–Ga hybridization and stabilizes the interfacial electronic structure, which is essential for maintaining low magnetic damping and robust PMA. Furthermore, if the interfacial anisotropy $K_i$ (> 0) is not negligible at such a chemically sharp and stabilized interface, it contributes additively to $K_u$. The effective anisotropy can thus be expressed as $K_{\text{eff}} = \frac{K_i}{t_{YIG}} + K_u - \frac{1}{2}\mu_0 M_s^2$, indicating that both preserved interface sharpness and strain-induced magnetoelastic effects cooperatively enhance PMA.

To validate this hypothesis, we perform microstructural and compositional analyses on YIG(10 nm)/GGG(111) and YIG(10 nm)/GSGG(111) samples using high-resolution transmission electron microscopy (HRTEM) and scanning TEM (STEM) with a JEOL JEM-2100F instrument. As shown in Figs. 3(a) and 3(d), both samples exhibit single-crystalline YIG layers and epitaxial interfaces between the YIG films and their respective substrates. To investigate interdiffusion across the YIG/substrate interfaces, energy-dispersive X-ray spectroscopy (EDS) mapping of constituent elements along the growth direction was performed. As shown in Figs. 3(b) and 3(c) for the YIG/GGG sample, and Figs. 3(e) and 3(f) for the YIG/GSGG sample, the substitution of Ga with Sc significantly suppresses the interdiffusion of Fe and Ga. This suppression reduces the probability of Fe atoms from the YIG film occupying Ga sites in the substrate and vice versa. Furthermore, the interdiffusion depth is reduced by a factor of 2 to 3, which is well consistent with the earlier discussion that the formation of the magnetic dead layer is significantly suppressed when YIG is grown on GSGG(111) (Fig. 2(g)). These results suggest that although GSGG substrates impose tensile strain on ultrathin YIG films, they less chemically promote interdiffusion, owing to the reduced $Ga^{3+}$ content and the presence of more stable, less mobile $Sc^{3+}$ ions. This highlights the importance of chemical and kinetic factors over purely strain-driven effects in governing interdiffusion and the formation of the magnetic dead layer.

**3.4. Ultralow magnetization damping in ultrathin YIG films at room temperature.** We now focus on the dynamic properties of ultrathin YIG films deposited on GGG(111) versus GSGG(111) substrates. Figures 4(a) and 4(b) show representative IP FMR spectra obtained at the fixed microwave frequency $f$ = 10 GHz and temperature $T$ = 300 K. Notably, all the FMR data are well-fitted using the $\mu_0 H$-derivative of a combination of symmetric and antisymmetric

Lorentzian functions[30,31],

$$\frac{d\chi''}{d\mu_0 H} \propto A \cdot \left[\frac{(\mu_0\Delta H_{HWHM})^2 \cdot (\mu_0 H - \mu_0 H_{res})}{[(\mu_0\Delta H_{HWHM})^2 + (\mu_0 H - \mu_0 H_{res})^2]^2}\right] + B \cdot \left[\frac{(\mu_0\Delta H_{HWHM}) \cdot (\mu_0 H - \mu_0 H_{res})^2}{[(\mu_0\Delta H_{HWHM})^2 + (\mu_0 H - \mu_0 H_{res})^2]^2}\right], \quad (3)$$

where $A$ ($B$) is the amplitude of the field derivative of the symmetric (antisymmetric) Lorentzian function. $\mu_0 H$ is the external DC magnetic field and $\mu_0\Delta H_{HWHM} = \frac{\sqrt{3}}{2}\mu_0\Delta H$ is the half-width-at-half-maximum (HWHM) of the imaginary part $\chi''$ of the magnetic susceptibility. This fitting procedure enables accurate extraction of the FMR linewidth $\mu_0\Delta H$ [related to the (effective) Gilbert damping $\alpha$] and the resonance field $\mu_0 H_{res}$ [associated with the (effective) saturation magnetization $\mu_0 M_s$].

Figures 4(c)-4(f) summarize the extracted $f$-dependent values of $\mu_0 H_{res}$ and $\mu_0\Delta H$ for various YIG films, from which the parameters $\alpha$, $\mu_0 M_s$, and the uniaxial magnetic anisotropy field $\mu_0 H_K$ can be estimated using the model developed earlier[40-43]:

$$\mu_0\Delta H(f) = \mu_0\Delta H_0 + \frac{4\pi\alpha f}{\sqrt{3}\gamma}, \quad (4)$$

$$\sin(2\theta_M) = \left(\frac{2\,\mu_0 H_{res}}{\mu_0 M_{eff}}\right) \cdot sin(\theta_H - \theta_M), \quad (5)$$

$$f = \frac{\gamma}{2\pi}\sqrt{\mu_0 H_1(\theta_H, \theta_M) \cdot \mu_0 H_2(\theta_H, \theta_M)}, \quad (6)$$

$$\mu_0 H_1(\theta_H, \theta_M) = \mu_0 H_{res} \cdot cos(\theta_H - \theta_M) - \mu_0 M_{eff} \cdot sin^2(\theta_M), \quad (7)$$

$$\mu_0 H_2(\theta_H, \theta_M) = \mu_0 H_{res} \cdot cos(\theta_H - \theta_M) + \mu_0 M_{eff} \cdot cos(2\theta_M). \quad (8)$$

Here, $\mu_0\Delta H_0$ is the zero-frequency line broadening due to long-range magnetic inhomogeneities, $\gamma = g_L\mu_B/\hbar$ is the gyromagnetic ratio ($1.84 \times 10^{11}$ $T^{-1}s^{-1}$), $g_L$ is the Landé g-factor (estimated to be 2.01 −2.04), $\mu_B$ is the Bohr magneton and $\hbar$ is Planck's constant divided by 2π. $\mu_0 M_{eff} = \mu_0 M_s - \mu_0 H_K$ is the effective magnetization and $\theta_M$ ($\theta_H$) is the OOP magnetization-angle of the YIG layer (magnetic-field-angle) relative to the interface plane. We consider the case in which the IP magnetization-angle $\phi_M$ of the YIG and the IP magnetic-field-angle $\phi_H$ are collinearly aligned ($\phi_M = \phi_H$), as relevant to our unpatterned YIG films.

We first note that successful linear fits to the $\mu_0 \Delta H(f)$ data using Eq. (4) [Figs. 4(d) and 4(f)] indicate the high quality of the YIG films and suggest that magnetization damping contribution from two-magnon scattering due to interface defects at room temperature (RT) is not significant[44,45]. For the 10-nm-thick (5-nm-thick) YIG film grown on GGG(111) substrates, we obtain $\mu_0 M_{eff}$ = 175 mT (157 mT) and $\alpha$ = 0.0016 (0.0085) [Figs. 4(c) and 4(e)]. These values are comparable to those reported in previous studies for samples with similar thicknesses[16,17,22,23]. Note that the deduced $\mu_0 M_{eff}$ is rather larger than the $\mu_0 M_s$ value from VSM data, suggesting the presence of nonzero $\mu_0 H_K < 0$[46]. This discrepancy is likely caused by growth-induced anisotropy. For the 3-nm-thick YIG film, no FMR signals are detected [Fig. 4(a)], which is consistent with $\delta \approx 2.3$ nm, indicating that the active thickness for ferrimagnetic ordering is less than one nanometer.

Importantly, when YIG films are grown on GSGG(111) substrates, FMR responses are observed even in the ultrathin regime down to 3 nm [Fig. 4(f)], accompanied by a clear magnetic anisotropy transition from IMA to PMA [Fig. 3(i) and 3(j)]. We estimate pronounced thickness-dependent $\mu_0 M_{eff}$ $(= \mu_0 M_s - \mu_0 H_K)$ values to be 28, –10, and –30 mT and ultralow $\alpha$ values to be 0.0006, 0.0008, and 0.001 for $t_{YIG}$ = 10, 5, and 3 nm, respectively. A positive $\mu_0 M_{eff}$ indicates that the magnetization easy axis lies IP whereas a negative $\mu_0 M_{eff}$ implies an OOP easy axis. Hence, combined static and dynamic analyses (Figs. 2 and 4) at RT confirm that our tensile-strained, ultrathin YIG films (≤ 10 nm) feature both tunable magnetic anisotropy with $\mu_0 H_K$ reaching up to 130 mT and ultralow damping constants of $\alpha \leq 10^{-3}$. These properties are attributed to tensile-strain-induced PMA and a reduced interfacial magnetic dead layer ($\delta \approx 0.7$ nm), driven by enhanced interfacial chemical stability in the YIG/GSGG(111) samples, as discussed above.

### 3.5. Significant reduction of cryogenic damping losses in chemically stabilized, tensile-

**strained YIG films.** Next, we investigate how the cryogenic damping losses of ultrathin YIG films depend on the substrate by measuring the $T$ evolution of the FMR spectra. Figure 5(a) shows typical IP FMR data obtained at the fixed $f$ of 10 GHz, from 300 K down to 10 K. In the case of YIG films grown on GGG(111) substrates [Fig. 5(a)], as $T$ decreases, $\mu_0\Delta H$ broadens greatly, $\mu_0 H_{\mathrm{res}}$ shifts to lower values, the FMR signal quickly diminishes. Below 150 K for the 10-nm-thick YIG film and below 150 K for the 5-nm-thick YIG film, the FMR signals become too weak to be detected using our coplanar-waveguide-based setup. These trends are qualitatively consistent with previous FMR studies on sputtered 15-nm-thick YIG thin films[16], where impurity relaxation[47], such as that caused by rare earth or $Fe^{2+}$ impurities, within the YIG film dominates damping losses at low $T$.

In contrast, tensile-strained YIG films grown on GSGG(111) substrates [Fig. 5(b)], exhibit substantially reduced magnetic damping at low $T$ compared to relaxed YIG films on GGG(111). With decreasing $T$, $\mu_0\Delta H$ broadens gradually, $\mu_0 H_{\mathrm{res}}$ shows a slight variation, and the FMR signal remains detectable at much lower $T$. Notably, for the 10-nm-thick YIG film [Fig. 5(b)], FMR spectra can be reliably measured down to 10 K [and even 2 K at the lowest, see Fig. 6(a) and Supplementary Section 2] using standard coplanar-waveguide FMR, without the need for a cavity-based system. Figures 5(c) and 5(d) summarize the $T$ dependence of $\mu_0\Delta H$ and $\mu_0 H_{\mathrm{res}}$, respectively, for the relaxed and tensile-strained YIG films. Overall, in addition to tunable magnetic anisotropy, the cryogenic FMR performance is markedly enhanced when YIG is grown on GSGG(111) compared to GGG(111).

**3.6. Estimation of low-temperature dynamic properties of tensile-strained YIG films.** For a quantitative analysis, we measure the FMR spectra as a function of $f$ at the fixed $T$ of ranging from 10 K down to 2 K [Fig. 6(a)] and determine the $\mu_0 H_{res}$ and $\mu_0\Delta H$ by fitting the spectra using a Lorentzian function based on Eq. (3) (Supplementary Section 2). Subsequently, the parameters $\mu_0 M_{eff}$ and $\alpha$ are extracted by fitting the $f$-dependent $\mu_0 H_{res}$ and $\mu_0 \Delta H$ data using Eqs. (4) and (6), respectively. We observe that $\mu_0 \Delta H$ scales quasi-linearly with $f$ in the low frequency regime down to 2 K and the $\mu_0 M_{eff}$ $(= \mu_0 M_s - \mu_0 H_K)$ value remains comparable to those at 300 K. The quasi-linear scaling of $\mu_0 \Delta H$ with $f$ up to about 45 GHz (Supplementary Section 3), sustained at least down to 50 K, indicates that two-magnon scattering is unlikely to play a major role in magnetic damping losses within this temperature range[44,45]. Moreover, the tensile-strain-induced $\mu_0 H_K$ is maintained throughout the cryogenic regime, persisting to temperatures as low as 2 K.

Perhaps, the most noteworthy dynamic aspect of the chemically stabilized, tensile-strained YIG is that although the extracted value of $\alpha$ = 0.011 (0.019) at 10 K (2 K) is approximately 18 (31) times larger than at 300 K, it still remains comparable to or even smaller than that of 3d transition metal ferromagnets of similar thickness[48,49]. Our result is also comparable to Y- or Bi- doped YIG films (relatively thicker, 28-36 nm)[50] that are lattice-matched and sputtered onto diamagnetic $Y_3Sc_2Ga_3O_{12}$ (YSGG) substrates, which exhibit remarkably improved cryogenic FMR properties (*i.e.* $\alpha = 0.002$ at 13 K), albeit without PMA. 30 nm. This highlights the unprecedented potential of Sc-substituted GGG substrates to suppress cryogenic damping losses in ultrathin YIG films while preserving the strain-induced PMA. A concise comparison of key parameters (e.g., thickness, substrate, magnetization damping at 300 K and 2–10 K, PMA strength, and magnetic dead-layer thickness) with previously reported studies[22,26,50,51] is provided in Supplementary Section 4.

We finally discuss how the paramagnetism of Gd in the GSGG substrate influences the low-$T$ magnetization dynamics of YIG films, particularly below 10 K. As demonstrated in recent studies of YIG films grown on garnet substrates[51-55], damping losses associated with impurity-related relaxation channels within the YIG layer are expected to freeze out at such low $T$[51,52]. Consequently, a persistent or enhanced FMR linewidth for $T \leq 10$ K suggests that substrate-induced paramagnetic losses start to play a dominant role[52-55]. A shift of the FMR resonance field toward higher values has been identified as an experimental signature of stray magnetic fields arising from the partially and inhomogeneously magnetized GGG substrate[54], which can artificially broaden the FMR linewidth and thereby increase the effective magnetization damping[52-55]. However, our tensile-strained ultrathin YIG do not exhibit such a resonance-field shift (Supplementary Section 5), indicating that a more microscopic exchange-coupling between Fe and Gd moments, rather than stray or dipolar fields, is likely responsible for the increased magnetization damping observed in our ultrathin YIG system. To further clarify this mechanism, a systematic comparative low-*T* ($\leq$ 10 K) FMR study of ultrathin ($\leq$ 10 nm) YIG films grown on GSGG and YSGG substrates will be necessary in the future.

## 4. CONCLUSIONS

We have demonstrated that ultrathin, tensile-strained YIG films grown by PLD on GSGG(111) substrates exhibit significantly reduced low-*T* magnetic damping compared to relaxed YIG films grown on lattice-matched GGG(111). Static magnetometry, broadband FMR measurement, and microstructural analysis together indicate that the superior dynamic performance of YIG on GSGG(111) arises from suppressed interdiffusion at the interface, enabled by favorable growth kinetics and enhanced chemical stability associated with the Sc-containing substrate. These findings underscore the critical role of chemical and kinetic factors in realizing few-nanometer-thick YIG film with negligible low-*T* damping losses and perpendicular magnetic anisotropy for high-performance, cryogenically operable spintronic and magnonic devices[31,56].

## ASSOCIATED CONTENT

**Supporting Information.** The Supporting Information is available free of charge at

https://pubs.acs.org/doi/XXX. Sections S1–S5 include the following: XRR analysis; FMR spectra measured below 10 K; broadband FMR measurements of relaxed and strained YIG films up to ~45 GHz; comparison of key parameters with previously reported studies; and the impact of the paramagnetism of GSGG substrates on low-temperature magnetization dynamics.

## ACKNOWLEDGMENTS

**Funding.** This work was supported by the Chung-Ang University Research Scholarship Grants (2024), the National Research Foundation (NRF) of Korea (Grant No. 2020R1A5A1016518), and the European Union through an ERC Advanced Grant (SUPERMINT, Project No. 101054860). Dong-Chan Kim gratefully acknowledges Hyeon Han (POSTECH) for initial guidance on the pulsed laser deposition growth of epitaxial ultrathin YIG films on GGG substrates. **Author Contributions.** K.-R.J. and S.S.P.P. conceived and designed the experiments. D.K. grew the ultrathin YIG films by pulsed laser deposition and performed X-ray diffraction structural analysis. Y.-C.L., J.-C.J., and J.Y. contributed to static magnetometry measurements. J.H.K. and S.-G.L. conducted broadband ferromagnetic resonance measurements, atomic force microscopy characterization, and transmission electron microscopy–based microstructural and compositional analyses under the supervision of K.-R.J. S.K., R.A., and T.T. performed cross-check FMR measurements. K.-R.J. and S.K. analyzed the FMR data. K.-R.J., S.K., and S.S.P.P. wrote the manuscript with contributions from all coauthors. **Competing Interests.** The authors declare no competing interests. **Data and Materials Availability.** All data required to evaluate the conclusions of this study are included in the article and/or the Supporting Information.

## FIGURE CAPTIONS

**FIG. 1.** (a) $\theta$–$2\theta$ X-ray diffraction (XRD) scans and (b) X-ray reflectivity (XRR) profiles for ultrathin YIG films with thicknesses $t_{YIG}$ = 3, 5, and 10 nm, grown on GGG(111) substrates. (c) and (d) Data equivalent to Figs. (a) and (b), but for YIG films with $t_{YIG}$ = 3, 5, and 10 nm, grown on GSGG(111) substrates. (e) Atomic force microscopy (AFM) scans for YIG films with $t_{YIG}$ = 3, 5, and 10 nm, respectively, grown on GGG(111). (f) Data equivalent to Fig. (e), but for YIG films with $t_{YIG}$ = 3, 5, and 10 nm, respectively, grown on GSGG(111) substrates.

**FIG. 2.** (a) In-plane (IP) and out-of-plane (OOP) magnetization $\mu_0 M$, normalized by their respective saturation magnetization $\mu_0 M_s$ values, versus external magnetic field $\mu_0 H$ for YIG films with $t_{YIG}$ = 3, 5, and 10 nm, respectively, grown on GGG(111) substrates. (b) Data equivalent to Fig. (a), but for YIG films with $t_{YIG}$ = 3, 5, and 10 nm, grown on GSGG(111) substrates. (c) Summary of the areal saturation magnetization $M_s \cdot t_{YIG}$ as a function of $t_{YIG}$.

**FIG. 3**. (a) High-resolution transmission electron microscopy (HRTEM) micrograph of YIG(10 nm) grown on GGG(111). (b) and (c) Spatial distribution of constituent elements obtained along the yellow arrow in (b) using spatially resolved energy-dispersive X-ray

spectroscopy (EDS). (d)-(f) Corresponding data for YIG(10 nm) grown on GSGG(111), equivalent to Figs. (a)-(c).

**FIG. 4.** (a) Typical in-plane ferromagnetic resonance (IP FMR) spectra for YIG films with $t_{\mathrm{YIG}}$ = 3, 5, and 10 nm, respectively, grown on GGG(111) substrates. These are performed at the fixed microwave frequency $f$ = 10 GHz and temperature $T$ = 300 K. (b) Data equivalent to Fig. (a), but for YIG films with $t_{\mathrm{YIG}}$ = 3, 5, and 10 nm, respectively, grown on GSGG(111) substrates. (c) Microwave frequency $f$ versus resonance field $\mu_0 H_{\mathrm{res}}$ for YIG films with $t_{\mathrm{YIG}}$ = 3, 5, and 10 nm, grown on GGG(111). (d) FMR linewidth $\mu_0\Delta H$ as a function of $f$ for YIG films with $t_{\mathrm{YIG}}$ = 3, 5, and 10 nm, grown on GGG(111). (e) and (f) Data equivalent to Figs. (c) and (d), but for YIG films with $t_{\mathrm{YIG}}$ = 3, 5, and 10 nm, grown on GSGG(111) substrates. Note that because the satellite peak observed in 10-nm-thick YIG of Fig. 4(b) exhibits a strong dependence on both magnetic field and frequency, we therefore attribute it most likely to a higher-order spin-wave mode rather than to a substrate contribution or a measurement artefact.

**FIG. 5.** (a) Representative IP FMR spectra measured at a fixed $f$ of 10 GHz, from 300 K down to 10 K, for YIG films with $t_{\mathrm{YIG}}$ = 3, 5, and 10 nm, respectively, grown on GGG(111) substrates. (b) Data equivalent to Fig. (a), but for YIG films with $t_{\mathrm{YIG}}$ = 3, 5, and 10 nm, respectively, grown on GSGG(111) substrates. (c) Summary of the FMR linewidth $\mu_0\Delta H$ as a function of temperature $T$ for YIG films with $t_{\mathrm{YIG}}$ = 3, 5, and 10 nm, grown on GGG(111). The inset shows the associated $T$ evolution of the resonance field $\mu_0\Delta H_{res}$. (d) Data equivalent to Fig. (c), but for YIG films with $t_{\mathrm{YIG}}$ = 3, 5, and 10 nm, grown on GSGG(111) substrates.

**FIG. 6.** (a) Representative $f$-dependent plots of $\mu_0 H_{\mathrm{res}}$ and $\mu_0\Delta H$ for the 10-nm-thick YIG film grown on a GSGG(111) substrate, measured at fixed temperatures of 10, 6, and 2 K over a frequency range of 3–10 GHz. The Gilbert damping parameter, $\alpha$, is extracted in the frequency range where $\mu_0\Delta H$ scales linearly with $f$. (b) Summary of the estimated $\alpha$ as a function of $T$ for YIG films grown on GGG(111) and GSGG(111), presented on a logarithmic scale for both axes. The results highlight the substrate-dependent influence on low-$T$ magnetic damping. A complete dataset, including FMR spectra below 10 K and broadband FMR measurements on relaxed and strained YIG films up to ~45 GHz from 300 K down to 10 K, is provided in the Supplementary Information.

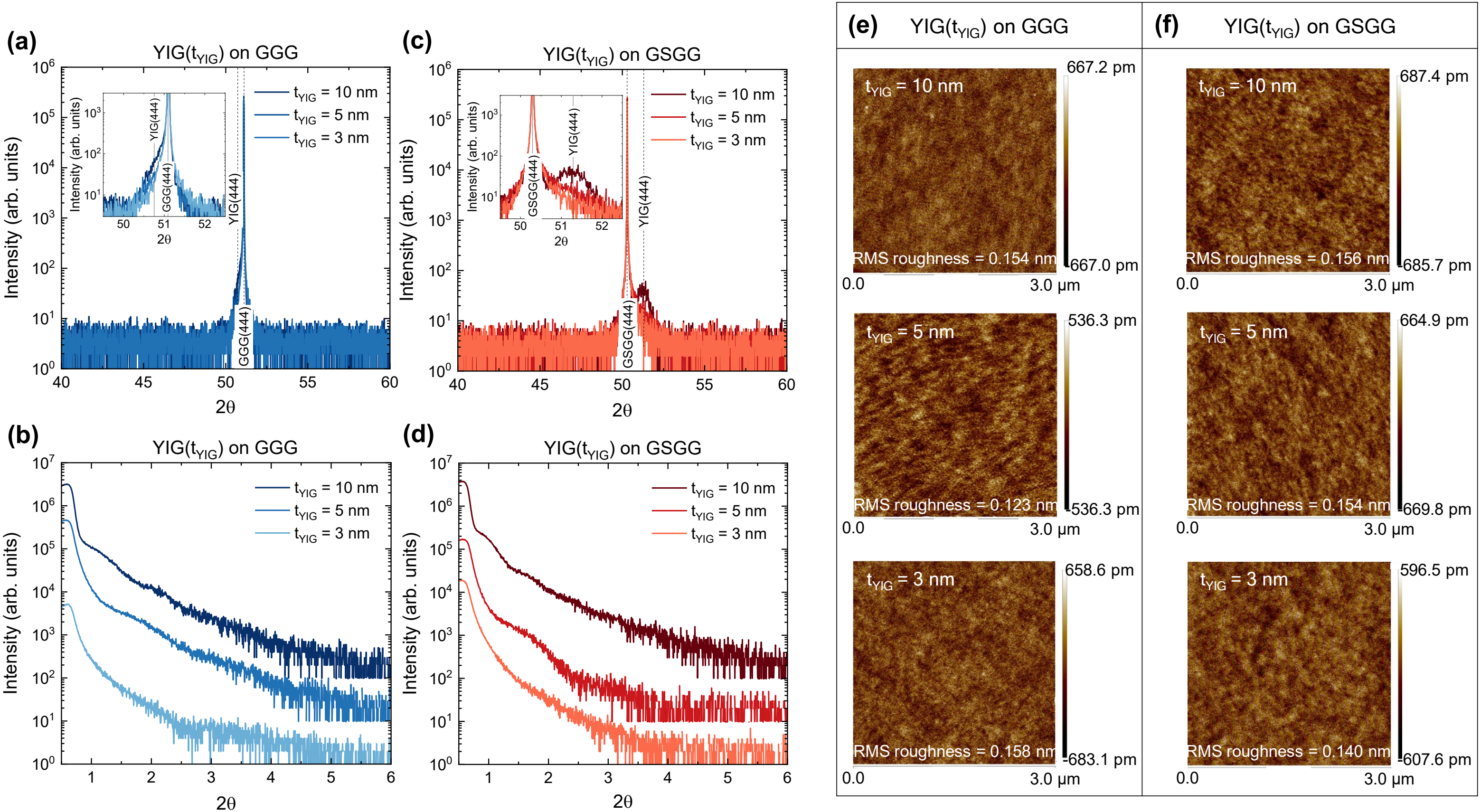

(a)
YIG(t_YIG) on GGG
t_YIG = 10 nm
t_YIG = 5 nm
t_YIG = 3 nm
YIG(444)
GGG(444)
Intensity (arb. units)
2θ
(b)
YIG(t_YIG) on GGG
(c)
YIG(t_YIG) on GSGG
GSGG(444)
(d)
YIG(t_YIG) on GSGG
(e) YIG(t_YIG) on GGG
(f) YIG(t_YIG) on GSGG
RMS roughness = 0.154 nm
RMS roughness = 0.123 nm
RMS roughness = 0.158 nm
RMS roughness = 0.156 nm
RMS roughness = 0.154 nm
RMS roughness = 0.140 nm
667.2 pm
667.0 pm
536.3 pm
536.3 pm
658.6 pm
683.1 pm
687.4 pm
685.7 pm
664.9 pm
669.8 pm
596.5 pm
607.6 pm
0.0
3.0 µm

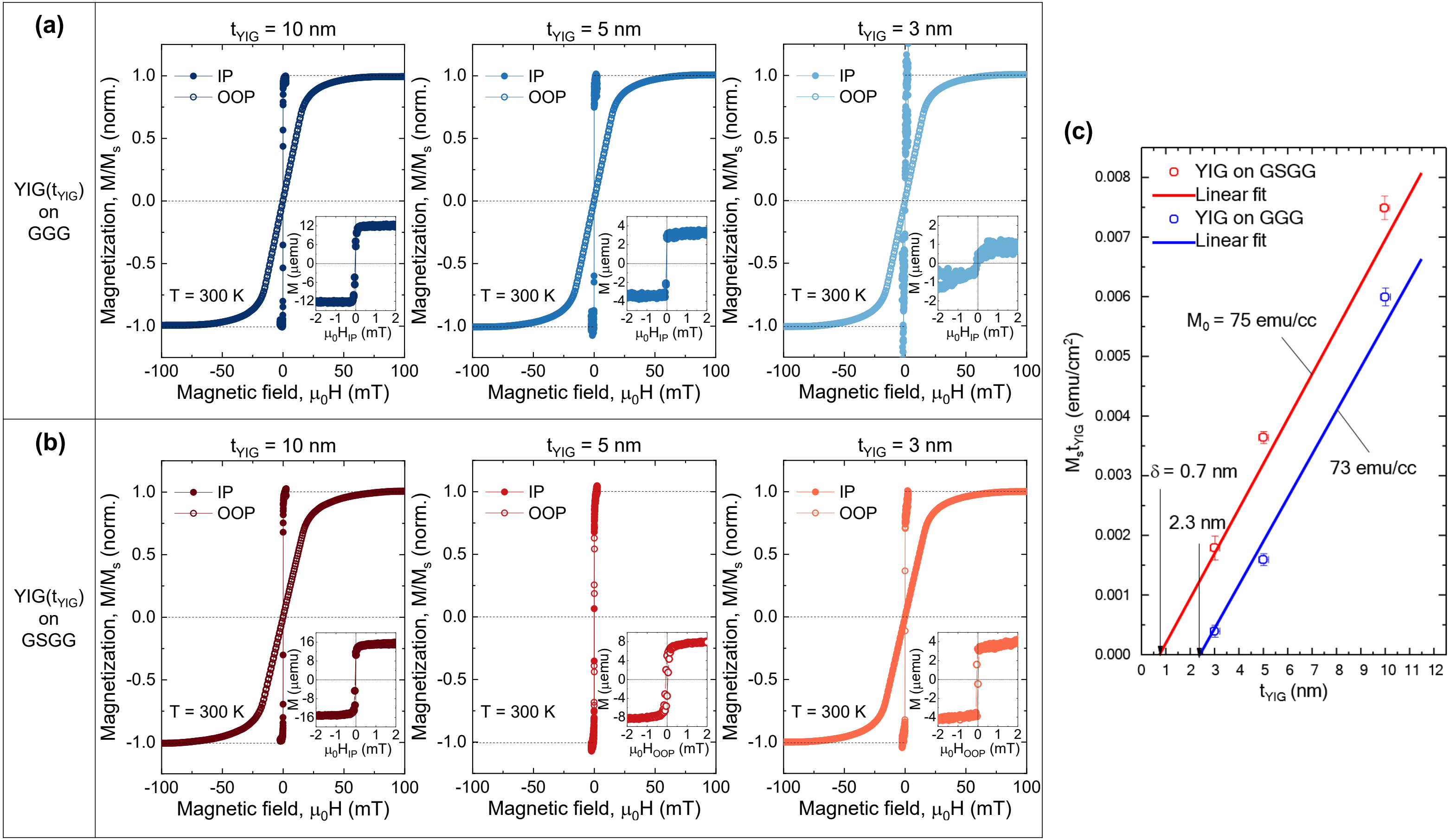

(a)
YIG($t_{YIG}$) on GGG
$t_{YIG}$ = 10 nm
$t_{YIG}$ = 5 nm
$t_{YIG}$ = 3 nm
IP
OOP
T = 300 K
Magnetization, $M/M_s$ (norm.)
Magnetic field, $\mu_0H$ (mT)
M (μemu)
$\mu_0H_{IP}$ (mT)
(b)
YIG($t_{YIG}$) on GSGG
$\mu_0H_{OOP}$ (mT)
(c)
YIG on GSGG
Linear fit
YIG on GGG
Linear fit
$M_0$ = 75 emu/cc
73 emu/cc
δ = 0.7 nm
2.3 nm
$M_st_{YIG}$ (emu/cm$^2$)
$t_{YIG}$ (nm)

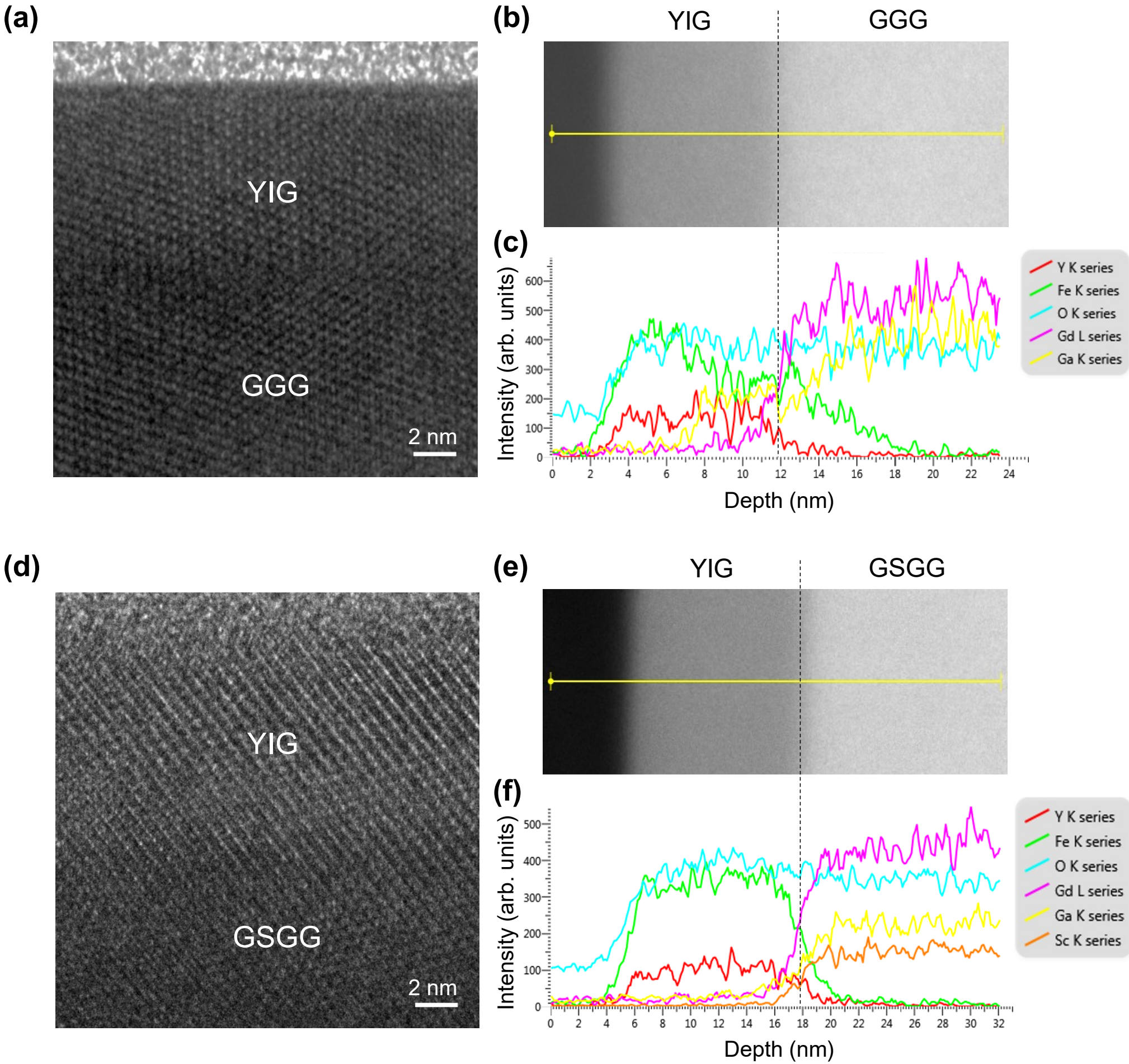

(a)
YIG
GGG
2 nm
(b)
YIG
GGG
(c)
Intensity (arb. units)
Depth (nm)
Y K series
Fe K series
O K series
Gd L series
Ga K series
(d)
YIG
GSGG
2 nm
(e)
YIG
GSGG
(f)
Intensity (arb. units)
Depth (nm)
Y K series
Fe K series
O K series
Gd L series
Ga K series
Sc K series

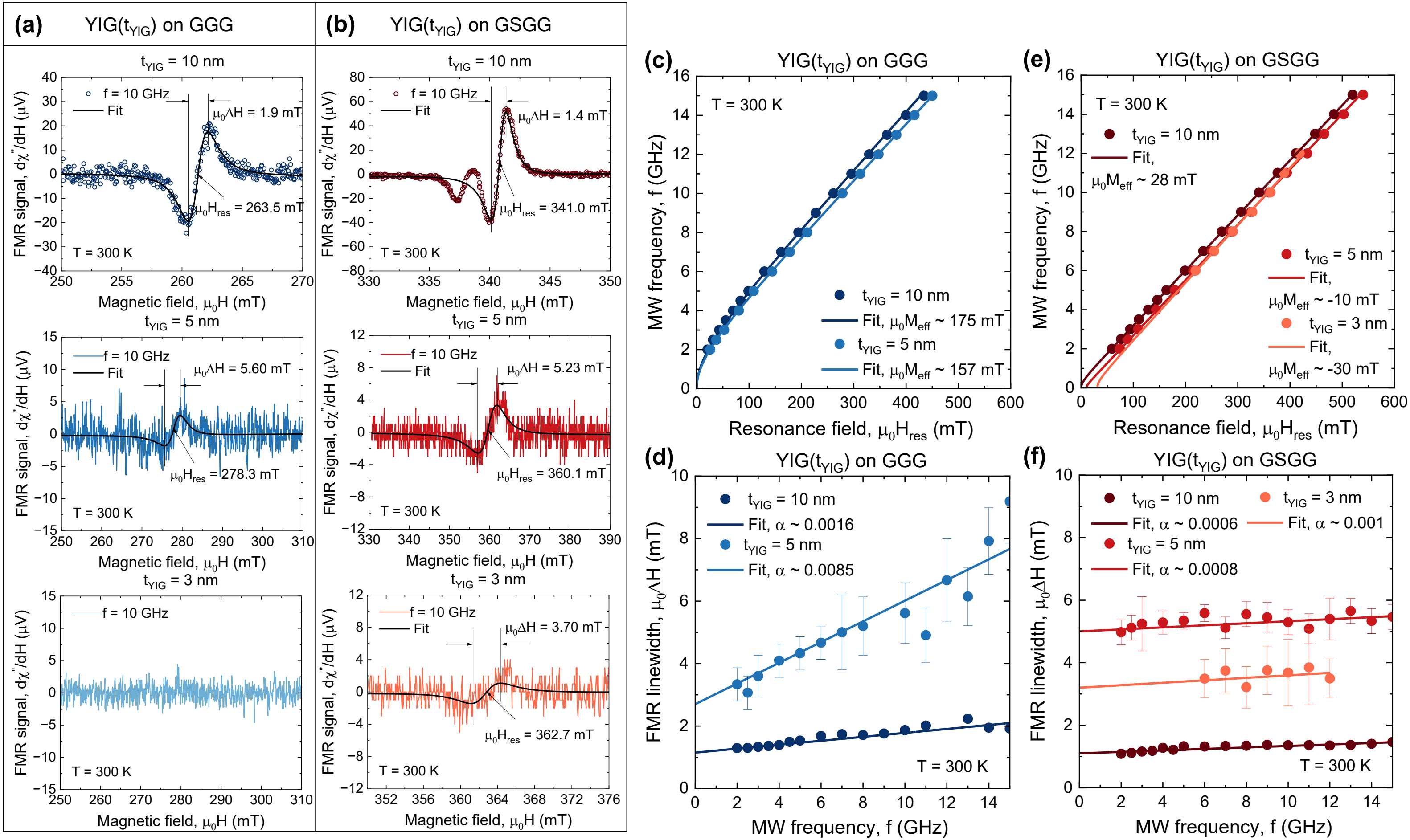

(a) YIG(t_YIG) on GGG
(b) YIG(t_YIG) on GSGG
t_YIG = 10 nm
f = 10 GHz
Fit
μ0ΔH = 1.9 mT
μ0Hres = 263.5 mT
T = 300 K
μ0ΔH = 1.4 mT
μ0Hres = 341.0 mT
t_YIG = 5 nm
μ0ΔH = 5.60 mT
μ0Hres = 278.3 mT
μ0ΔH = 5.23 mT
μ0Hres = 360.1 mT
t_YIG = 3 nm
μ0ΔH = 3.70 mT
μ0Hres = 362.7 mT
FMR signal, dχ''/dH (μV)
Magnetic field, μ0H (mT)
(c) YIG(t_YIG) on GGG
t_YIG = 10 nm
Fit, μ0Meff ~ 175 mT
t_YIG = 5 nm
Fit, μ0Meff ~ 157 mT
MW frequency, f (GHz)
Resonance field, μ0Hres (mT)
(e) YIG(t_YIG) on GSGG
t_YIG = 10 nm
Fit, μ0Meff ~ 28 mT
t_YIG = 5 nm
Fit, μ0Meff ~ -10 mT
t_YIG = 3 nm
Fit, μ0Meff ~ -30 mT
(d) YIG(t_YIG) on GGG
Fit, α ~ 0.0016
Fit, α ~ 0.0085
FMR linewidth, μ0ΔH (mT)
MW frequency, f (GHz)
(f) YIG(t_YIG) on GSGG
Fit, α ~ 0.0006
Fit, α ~ 0.001
Fit, α ~ 0.0008

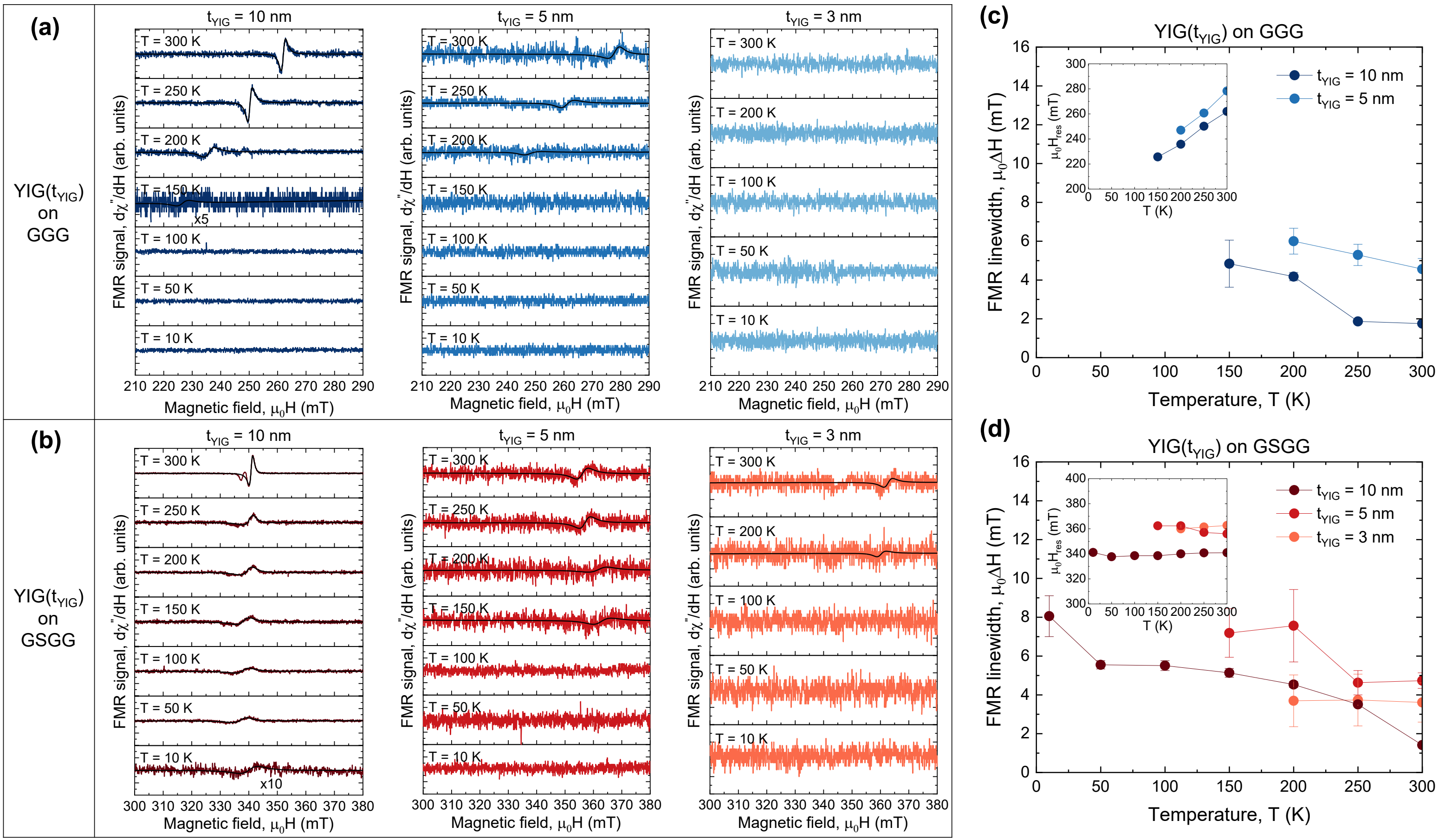

(a)
YIG($t_{YIG}$) on GGG
$t_{YIG}$ = 10 nm
$t_{YIG}$ = 5 nm
$t_{YIG}$ = 3 nm
T = 300 K
T = 250 K
T = 200 K
T = 150 K
T = 100 K
T = 50 K
T = 10 K
x5
FMR signal, dχ"/dH (arb. units)
Magnetic field, $\mu_0 H$ (mT)
(b)
YIG($t_{YIG}$) on GSGG
x10
(c)
YIG($t_{YIG}$) on GGG
FMR linewidth, $\mu_0 \Delta H$ (mT)
Temperature, T (K)
$\mu_0 H_{res}$ (mT)
T (K)
(d)
YIG($t_{YIG}$) on GSGG

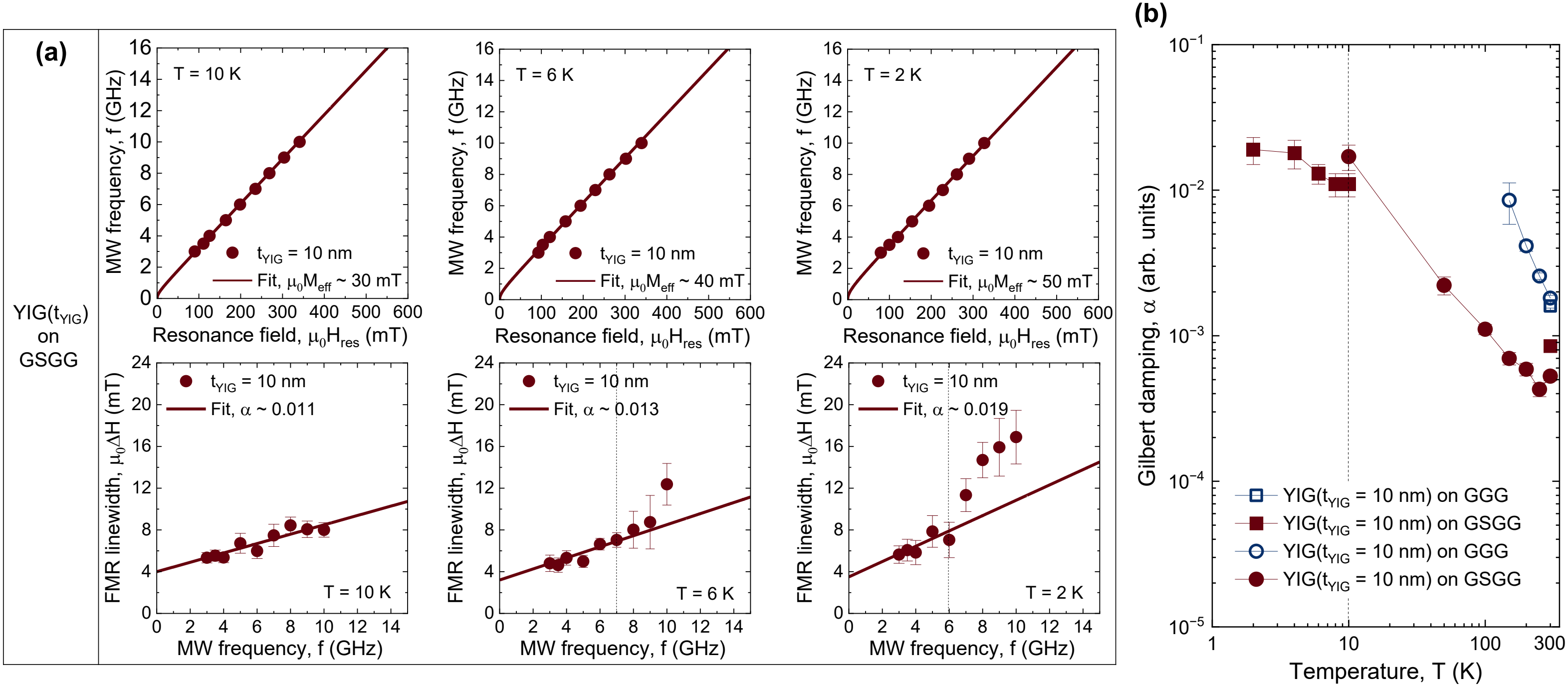
(a)
YIG($t_{YIG}$) on GSGG
T = 10 K
T = 6 K
T = 2 K
$t_{YIG}$ = 10 nm
Fit, $\mu_0 M_{eff}$ ~ 30 mT
Fit, $\mu_0 M_{eff}$ ~ 40 mT
Fit, $\mu_0 M_{eff}$ ~ 50 mT
MW frequency, f (GHz)
Resonance field, $\mu_0 H_{res}$ (mT)
Fit, α ~ 0.011
Fit, α ~ 0.013
Fit, α ~ 0.019
FMR linewidth, $\mu_0 \Delta H$ (mT)
(b)
Gilbert damping, α (arb. units)
Temperature, T (K)
YIG($t_{YIG}$ = 10 nm) on GGG
YIG($t_{YIG}$ = 10 nm) on GSGG
YIG($t_{YIG}$ = 10 nm) on GGG
YIG($t_{YIG}$ = 10 nm) on GSGG